4# Numerical Study on Human Brain Cortical Electrostimulation Assessment During Uniform Magnetic Field Exposure at Intermediate Frequencies

Jose Gomez-Tames, Thomas Tarnaud, Wout Joseph, Emmeric Tanghe*Abstract Objectives:* Permissible limits have been established by international guidelines and standards for human protection to electromagnetic field exposure to prevent adverse health effects stemming from electrostimulation in the most sensitive body part. That is the peripheral nervous system (PNS) in the intermediate frequency range (300 Hz to 100 kHz) and the central nervous system (CNS) at lower frequencies. However, there is a need to reevaluate protection limits against CNS electrostimulation in the intermediate frequency range, considering the importance of brain tissues during electromagnetic head exposure. This study aims to derive the level of CNS cortical stimulation to evaluate compliance with existing protection limits. *Method:* Multi-scale computation modelling was used to evaluate neuron stimulation thresholds by integrating individual neurons into realistic anatomical head models. Five different excitable membrane models within the motor cortex were examined across three human head models, providing the most comprehensive and extensive evaluation to date. *Results:* Current protection limits are confirmed as conservative, with non-compliance observed in only 0.02% and 2.4% of axons under clamped and sealed boundary conditions, respectively. The study highlights significant intersubject variability (up to 600% mean threshold) and clarifies the influence of neural excitation models on permissible level assessments. *Conclusion:* Current electric field limits are conservative for CNS electrostimulation in the intermediate frequencies range, but the margin of safety decreases at higher frequencies, warranting further evaluation. *Impact:* The study's findings and methodology contribute to the rationale and provide valuable insights for revising electromagnetic safety exposure guidelines.

*Keywords* **— Dosimetry, Intermediate frequency, Human safety, Multi-scale modeling, Nerve model, Protection limits, Standardization**Note: There should no nonstandard abbreviations, acknowledgments of support, references or footnotes in in the abstract.

## I. INTRODUCTION

Limiting potential adverse health effects of humans to non-ionizing electromagnetic field exposure has motivated the development of international guidelines and standards (International Commission on Non-Ionizing Radiation Protection or ICNIRP [1] and the Institute of Electrical and Electronics Engineers International Committee on Electromagnetic Safety or IEEE ICES [2]). Exposure limits have been established to prevent adverse health effects based on electrostimulation and thermal effects. The electrostimulation effect is dominant up to 100 kHz for continuous exposure (and up to 5–10 MHz for brief pulse exposures), while the thermal effect becomes dominant at frequencies higher than 100 kHz for continuous exposure due to energy absorption from electromagnetic fields.

Regarding protection from adverse effects based on electrostimulation, the guidelines aim to protect the most sensitive region of the body. This would include adverse effects on the peripheral nervous system due to electrostimulation of axons (PNS) for frequencies above ~400 in ICNIRP guidelines and ~750 Hz in IEEE to 100 kHz. Below the lower frequencies, the guidelines protect against adverse central nervous system stimulation (CNS) related to adverse synaptic effects and are based on phosphenes as a surrogate.

The exposure limits, incorporating safety margins, refer to limits of the internal and external allowable quantities. In the context of electrostimulation, the internal allowable quantity is the internal electric field induced within the exposed individual's body. This quantity is termed dosimetric reference limits (DRL) in IEEE and basic restriction (BR) in ICNIRP. The external allowable quantity pertains to the external magnetic/electric field strength in the absence of human presence, termed exposure reference levels (ERL) in IEEE and reference level (RL) in ICNIRP. ICNIRP

T. T. was a postdoctoral fellow of the FWO‑V (research foundation Flanders, 1230222N). J. G-T was supported by a JSPS Grant-in-Aid for Scientific Research, JSPS KAKENHI Grant (23K25176).J. G-T is with the Graduate School of Science and Engineering, Chiba University, Chiba 2638522, Japan, and also with the Center for Frontier Medical Engineering, Chiba University, Chiba 2638522, (e-mail: jgomez@chiba-u.jp). T. T, W. J, and E. T are with Department of Information Technology (INTEC), Ghent University-imec, Ghent 9052. T. T is also with Department of Head and Skin, Ghent University, Ghent 90001 delineates its restrictions based on two categories: one applicable to the general public and another to occupational exposure, where the latter permits higher thresholds (less conservative). Meanwhile, IEEE classifies exposure scenarios as either restricted or unrestricted environments.

In the IEEE standard, DRLs for the peripheral nervous system were derived from an internal electric field threshold estimated from a 20 μm fiber model (SENN: spatially extended nonlinear node) excited by a uniform electric field [3], consistent with experimental thresholds [4]. Reduction factors were then applied to these thresholds to obtain the DRLs. In the ICNIRP guidelines for the peripheral nervous system, the internal electric field threshold was based on dosimetry analysis of experimental measurements to which the reduction factor is applied to obtain the BRs.

Permissible levels in the intermediate frequencies (300 Hz to 100 kHz) in the case of CNS tissue instead of PNS need to be revised, considering the importance of brain tissues during electromagnetic head exposure. Derivation of the threshold level for the CNS (direct axonal stimulation; no synaptic effect) in the intermediate frequencies for various head models and axon models would be helpful for the scientific soundness of the guidelines/standards.

To understand external field stimulation and derive thresholds scientifically, multi-scale simulations by combining electric field analysis in biological tissues (mm-level) and neuron models (micro-meter level) are necessary, as outlined in the research agenda of IEEE ICES [5]. Additionally, ICNIRP recommends conducting excitation modeling and threshold assessments to improve the accuracy of the restrictions [6].

Computational dosimetry becomes an essential tool for estimating induced physical quantities. Recently, incorporating individual neurons into realistic head models facilitates investigating neuron stimulation thresholds[7]. This addresses the problem of evaluating electrostimulation thresholds for electromagnetic safety [8], [9]. In addition, experimental measurement of pain threshold is limited across the intermediate frequency range [10], and in vivo electric field measurements are limited.

A working group within the IEEE ICES Technical Committee 95 Subcommittee 6, led by the authors, was established to explore electrostimulation thresholds in the brain. The consistency of the multi-scale computations was investigated, and the conservativeness of permissible field strengths of international guidelines at intermediate frequencies was evaluated [11]. However, a comprehensive assessment requires considering variability by accounting for different excitation neural membrane models and anatomical differences in head anatomy that may significantly impact compliance with the protection limits, which is missing in the literature.

This study aims to investigate brain cortical stimulation thresholds, considering various classical neural models and individual differences in head anatomy, using a multi-scale computational approach for uniform magnetic field exposure at intermediate frequencies. Moreover, the conservativeness of protection limits in international guidelines is investigated. Importantly, the results and methodology of this study could contribute to the rationale and serve as input for the revision or updates of the intermediate frequency exposure guidelines.

## II. MODELS AND METHODS

### A. Human Head Model

Three realistic head models (all male, 41 years old, healthy) were created from freely available magnetic resonance image (MRI) data. The head models were voxelized with a resolution of 0.5 mm and consisted of 12 tissues/body fluids, as shown in [12].

### B. Electromagnetic Computational Method

At the intermediate frequencies, the magneto-quasistatic approximation applies to the computation of the internal electric field in biological tissues. The electric and external magnetic fields are decoupled and treating the exposure to these fields separately is possible. Also, permittivity can be neglected; only tissue conductivity is considered, as the displacement currents are more than one order of magnitude smaller than the conduction current [13]. Thus, the induced scalar electric potential ϕ is given by the following equation:

$$\nabla \cdot \sigma \nabla \phi = -\nabla \cdot \sigma \frac{\partial \mathbf{A}}{\partial t}, \quad (1)$$

where $A$ and $\sigma$ denote the magnetic vector potential of the applied (external) magnetic field and tissue conductivity, respectively. When (1) is solved, the internal electric field $\mathbf{E}$ is calculated: $\mathbf{E} = -\nabla \phi - j\omega \mathbf{A}_0$.

Equation (1) was solved numerically by the scalar potential finite difference method [14]. Tissue conductivities were assumed linear and isotropic and were determined using the fourth-order Cole-Cole model [15] at the frequencies corresponding to different exposure scenarios.

### C. Neuronal Compartment Computational Models

We calculated stimulation thresholds of thick pyramidal axons extending from the hand motor cortex (Fig. 1A). Stimulation thresholds are identified when the induced electric field elicits action potentials in the cortical axon models. The axon model from [3] is utilized, representing myelinated axons comprising internodes (segments covered by myelin sheaths) interspersed with nodes of Ranvier (short segments with a high ionic channel density) [16]. At the nodes of Ranvier, the ionic membrane current is formulated based on five classical models that follow a conductance-based voltage-gated model (see Table 1) . These classical membrane models were the five earliest Hodgkin-Huxley type models obtained from patch clamp data. They were chosen for this study, because they are used frequently in computational neuroscience or neural engineering studies [17], e.g., for the modelling of the human cochlear nerve and conduction block [[17], [18]]). In particular, the Frankenhaeuser-Huxley model (FH) has been used in the IEEE ICES C95.6 and C95.1 standards [2] and in the ICNIRP 2010 guideline [5] To make the Hodgkin-Huxley (HH) model compatible with the myelinated axon model, the nodal conductivity and capacitance values were scaled, as in [19]. At the internodes, the membrane current was modelled by the passive conductance multiplied by the membrane potential.

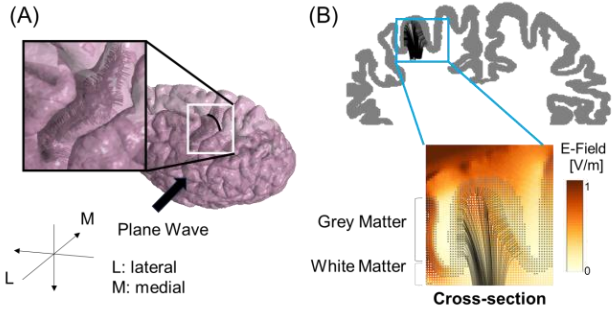

Fig. 1. (A) Cortical axons are placed on the hand motor area (hand knob). (B) Cross-section plane showing the distribution of the axons and local electric fields induced by uniform exposure. The arrow is the direction of the incident plane electromagnetic wave.

TABLE I
MEMBRANE MODELS APPLIED TO THE AXON MODEL

| Model | Experiment |
|---|---|
| Hodgkin-Huxley (**HH**) | Squid axon |
| Frankenhaeuser-Huxley (**FH**) | Frog node |
| Chiu-Ritchie-Rogart-Stagg-Sweeney (**CRRSS**) | Rabbit node |
| Schwarz-Eikhof (**SE**) | Rat node |
| Schwarz-Reid-Bostock (**SRB**) | Human Nerve |

Neural activation is driven by the extracellular potential along the axons (Fig. 1B). The extracellular potential is estimated as a local linear integral of the internal electric field along the axon trajectory (resulting in a "quasipotential", because the electric field is not conservative in general. The elicitation of an action potential is considered when the membrane potential is depolarized and exceeds 50 mV, with either propagation over at least four consecutive nodes of Ranvier in either direction, or symmetrical propagation in both the ortho- and antidromic directions over two nodes of Ranvier [20]. Thus, the threshold can be expressed as the required internal electric field (or external magnetic field strength) to elicit an action potential.

Two boundary conditions were selected to approximate a terminating axon without considering other parts of the neuron (soma, axon collaterals, and dendritic tree). The sealed boundary condition is normally used as boundary conditions at the terminations of axon collaterals (e.g., [21]). The voltage clamp boundary condition considers that the axon end does not exist by assuming a fixed value equal to the resting potential of the node of Ranvier. This boundary condition is motivated by the observation that, in reality, there would be a connection with the soma instead of this termination. Here, voltage clamp conditions might be a good first approximation for the axial current running from the axon terminal towards the soma.

*D. Exposures Scenarios*

We explored the influence of uniform exposure (300 Hz, 1 kHz, 10 kHz, 100 kHz) on neural activation of cortical axons (5 membrane models) embedded in three head models. We direct a uniform magnetic field (vector potential A in equation 1 is defined such that magnetic field is constant) from the lateral to medial direction with continuous sinusoidal pulses (Figure 1A). We opted for this direction due to its lower thresholds of the axons projecting from the putative hand motor area compared to the two other conventional uniform magnetic field exposure directions (anterior-posterior and superior-inferior) [11].

Fast-conducting thickly myelinated pyramidal tract axons (Betz cell's axon) were considered for the hand motor area (20 μm in diameter [22]). An average of approximately ten thousand neural axons are projected from the hand motor cortex of each head model [23], as illustrated in Figure 1B. These axons traverse the gray-white matter boundary nearly perpendicular, ultimately forming the pyramidal tracts. Considering the influence of the electric field on axon bending, we integrated the bending of pyramidal cell axons into our model for axonal projection from the gyral wall and crown. These axon pathways were situated within the putative hand knob region of the primary motor cortex, approximately 2.5 mm beneath the cortical surface. A detailed explanation of how we generated pyramidal axon models can be found in [23].

*E. Data Analysis*

Post-processing techniques are implemented to effectively mitigate outliers in electric field values inherent in utilizing voxelized anatomical models, where curved boundaries are approximated with a stair-step method. The 99th percentile field strength in the brain cortex is used to mitigate computational artifacts in the standards for scenarios of uniform exposure [24]. For a specific tissue, the 99th percentile value of the electric field is the relevant value to be compared with the DRLs and BRs.

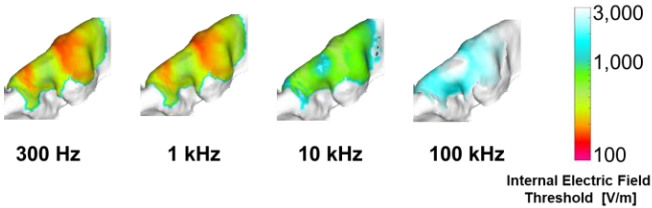

Fig. 2. Spatial distribution of cortical activation thresholds at different frequencies on the hand motor area for one subject (7,358 cortical axons) for FH voltage clamped.

## III. NUMERICAL RESULTS

### A. Electric field distribution and neural activation

Figure 2 shows activation maps at frequencies from 300 Hz to 100 kHz depicting the electric field threshold for each axon projecting from the cortical surface of the hand knob. Internal electric field thresholds are observed to increase at higher frequencies (35 and 8.3-fold increase over the range 300 Hz to 100 kHz of the mean and minimal thresholds, respectively). Also, despite some variability in the proportion of electrical conductivity among tissues at different frequencies, the shape of the activation maps remains consistent across the frequencies investigated. These observations were found to be consistent across all membrane neural models.

### B. Permissible Field Strength

Figures 3 and 4 show that the stimulation thresholds increase with frequency, irrespective of the neuronal membrane and anatomical head models. In Figure 3, the stimulation thresholds of all axons show the conservativeness of all the protection limits in the clamped terminal condition, except for 0.3% of the CRRSS axons in subject 2 (S2) exhibiting thresholds below the limits at 100 kHz in the restricted environment (representing only 0.02% of all axons in the clamped condition at 100 kHz). Figure 4 shows that all axons comply with the limits for frequencies ranging from 300 Hz to 10 kHz for the sealed terminal. Also, all axons are compliant at 100 kHz, except for 4%-18% (CRRSS, S1-S3) and 5% (FH, S2) in the IEEE restricted environment (representing only 2.4% of all axons in the sealed condition).

Figures 3 and 4 show that the most conservative limits are

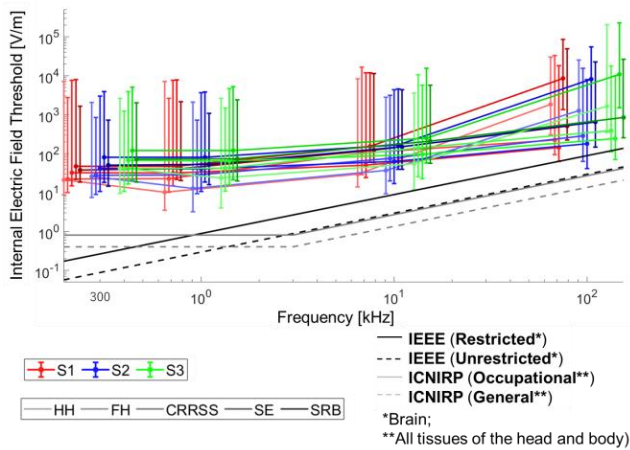

Fig. 4. Excitation thresholds for uniform exposure compared with the permissible internal limits of ICNIRP guidelines and the IEEE safety standard. The threshold is estimated using a sealed terminal condition. Excitation thresholds are plotted for 300 Hz, 1 kHz, 10 kHz and 100 kHz: horizontal offsets between different subjects and membrane models are added for visualization. **Abbreviations:** HH: Hodgkin-Huxley, FH: Frankenhaeuser-Huxley, CRRSS: Chiu-Ritchie-Rogart-Stagg-Sweeney, SE: Schwarz-Eikhof, SRB: Schwarz-Reid-Bostock.

presented for lower-end frequencies (300 Hz −1 kHz). In the voltage-clamped terminal condition, there is a 22.5-fold gap between the minimum axon threshold and the unrestricted limit at 1 kHz (S2, HH model). The minimum gap is reduced to 5.2 and 2.3 times at 10 and 100 kHz, respectively. The gap is smaller in the sealed terminal condition to 10.9 times, 3.1 times, and 1.3 times at 300 Hz −1 kHz, 10 kHz, and 100 kHz, respectively. There is also an important difference between the membrane models. For lower-end frequencies (300 Hz −1 kHz), the FH and HH models present the minimum thresholds and are closer to the limits (sealed or clamped terminals). For the middle frequency (10 kHz), the HH, FH, and CRRSS models present smaller thresholds. For the end frequency (100 kHz), FH and CRRSS models present the minimum thresholds. The SE model shows the highest minimum thresholds for all frequencies. Threshold computation implemented in sealed end terminals reduces the threshold compared to clamped terminals by a factor of

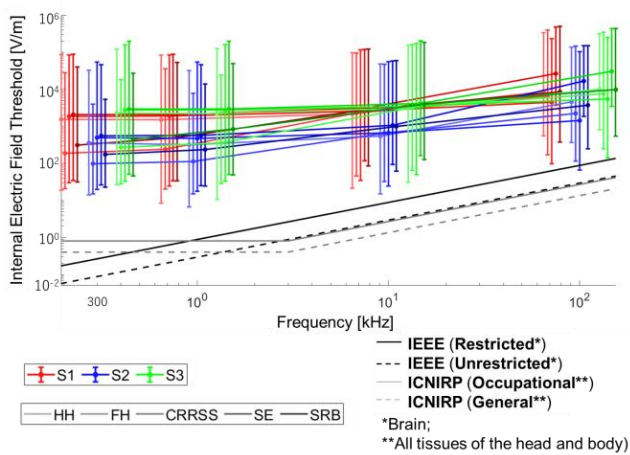

Fig. 3. Excitation thresholds for uniform exposure compared with permissible internal limits of ICNIRP guidelines and IEEE safety standards. The threshold is estimated using the clamped terminal condition. Excitation thresholds are plotted for 300 Hz, 1 kHz, 10 kHz and 100 kHz: horizontal offsets between different subjects and membrane models are added for visualization. Abbreviations: HH: Hodgkin-Huxley, FH: Frankenhaeuser-Huxley, CRRSS: Chiu-Ritchie-Rogart-Stagg-Sweeney, SE: Schwarz-Eikhof, SRB: Schwarz-Reid-Bostock.

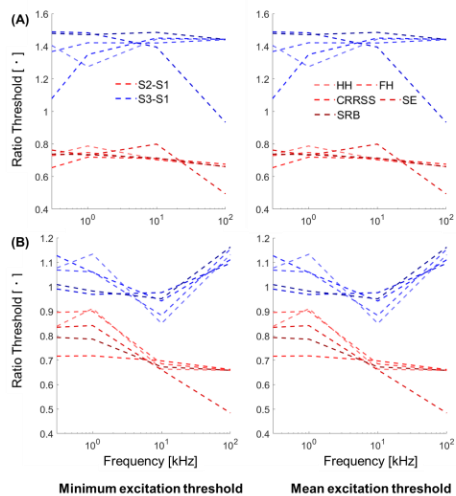

Fig. 5. Ratio of threshold variations among three head models using one as common reference (S1) for (a) clamped and (b) sealed terminal conditions

56.7 to 106.

*C. Individual difference*

Figure 5 compares excitation thresholds among the three head models across all conditions. The ratios of mean and minimum thresholds were calculated with one head model serving as the common reference (e.g., $S_3/S_1$ and $S_2/S_1$). In voltage-clamped conditions (Figure 5A), the variations among mean ratio thresholds range from 0.2 to 1.7, while in sealed conditions (Figure 5B), they range from 0.7 to 2.6. This represents variations of up to 127% in the minimum threshold among the head models for a fixed membrane model, as well as boundary condition and frequency. The variation among average thresholds exceeds that among minimum thresholds, reaching up to 600%. We observed a larger intersubject variation of the mean threshold ratio in the voltage-clamped terminal than the sealed terminal boundary condition across different frequencies and membrane models (a 3.6-fold increase of the variation on average between both boundary conditions).

## IV. DISCUSSION

This study examined CNS thresholds using axonal models projecting from the brain cortex to evaluate compliance with international electromagnetic exposure guidelines. IEEE Std C95.1-2019 specifies ERLs for whole-body exposure to prevent exceeding a DRL in the most sensitive regions. This corresponds to the PNS in the intermediate frequency. However, given the importance of CNS protection, it is necessary to revisit protection limits in this frequency range by incorporating axonal trajectories, anatomically realistic head models, and detailed membrane dynamics. Applying recent methodologies to investigate CNS thresholds will enhance the scientific basis of these standards and improve understanding of CNS exposure limits.

*A. Conservativeness*

Conservativeness is investigated considering various scenarios: five membrane models, an average of approximately ten thousand axonal fibers per head model, three head models, terminal conditions (sealed end or voltage clamped), and four frequencies (0.3, 1, 10, 100 kHz). In total, we conducted more than 1.2 million exposure scenarios to evaluate the conservativeness of the limits. For the general public (unrestricted environment), conservativeness was confirmed for the ICNIRP and IEEE limits in all conditions. For occupational exposure (restricted environment), a marginal number of axons (0.02% among all voltage clamp boundary condition simulations) did not comply with the limits at 100 kHz. More non-compliant axons (2.4% among all conditions) were observed in the restricted environment levels at 100 kHz for sealed terminals.

*B. Terminal boundary conditions*

The selection of the terminal boundary conditions influenced the stimulation threshold. Sealed ends reduced the threshold compared to voltage-clamped conditions with a factor of approximately 5 to 100. Voltage-clamped axons are generally more difficult to excite, making sealed terminals more conservative for challenging the limits. However, cortical stimulation thresholds under clamped terminal conditions at 1 kHz are closer to the 50 V/m threshold (the lower tolerance interval of the active motor threshold) observed in transcranial magnetic stimulation than those under sealed terminals [8]. This suggests that voltage-clamped conditions better reflect a realistic scenario. It is important to note that we used a conservative assumption of large fiber thickness (20 μm diameter in the case). For more common values of 1-5 μm [21], the thresholds are expected to increase by one to two orders of magnitudes. Under these conditions, the sealed ends may align more closely with experimentally observed thresholds, such as in a multi-scale modeling study of transcranial magnetic stimulation that used Blue Brain Project neural cells [21]. In this context, sealed thresholds would approach those of clamped conditions for a 20 μm [25]. Thus, voltage-clamped conditions remain suitable for comparison with protection limits. In the case of ICNIRP, BRs apply a five-fold safety margin for occupational environments and a 10-fold reduction for the general public. In the case of IEEE, the DRLs apply safety factors of three and nine are applied to body parts in unrestricted and restricted environments in the frequency range investigated in this study. Mean thresholds complied with the margins in all conditions. In the case of minimum thresholds, the margin was reduced at higher frequencies. The margins are satisfied with the minimum estimated cortical thresholds except at 100 kHz, in which the limits of the restricted environment were near the minimum thresholds.

*C. Limitations and future work*

First, an important consideration is that the excitation threshold strongly depends on how neuronal excitation is defined, especially at higher frequencies. The action potential is defined as depolarization over $V_T = 50$ mV and imposed conditions on action potential propagation over consecutive nodes of Ranvier. Varying depolarization threshold $V_T$ or AP propagation conditions are expected to change the excitation thresholds. At higher frequencies, the typical action potential shape can be lost by fast membrane voltage oscillations occurring simultaneously at subsequent nodes. If these oscillation amplitudes are sufficient, they are considered action potentials under our conditions, potentially causing false positives. Proper identification of these could reduce non-compliant axons. Second, as in the rationale of the safety guidelines/standard, we used neuronal excitation of an axon model constrained by boundary conditions to assess the conservativeness of safety guidelines. However, subthreshold electromagnetic fields that, below the action potential threshold, are still capable of altering brain function [26]. Furthermore, morphologically realistic models of human cortical cells, incorporating soma, dendritic trees, and axon collaterals, could enhance simulation accuracy. Future work should clarify subthreshold effects and the role of morphological realism in safety assessments. A source of uncertainty arises from electric conductivity variations (anisotropy, uniformity, age dependency) [27], [28]. Furthermore, this study highlights the importance of considering various head models, because of the significant intersubject variability observed in three head models (Fig. 5). Future research should explore larger datasets and improved volume conductor models, such as diffusion tensor imaging MRI-based anisotropic conductivities, for more accurate assessments of safety guideline conservativeness.

## V. CONCLUSIONS

We determined that the electric field limits established in both guidelines are conservative with the internal electric field needed for CNS stimulation in the intermediate frequency range, although the margin of conservativeness is reduced at higher frequencies when considering the strictest evaluations. For example, for voltage clamp conditions we observed 22.5 and 2.3 fold gaps between the lowest threshold and the unrestricted environment ICES guidelines at low-end (300 Hz – 1 kHz) and high-end (10 kHz - 100 kHz) frequencies, respectively. An extensive comparison considering more than 1.2 million scenarios from different nerve models, implementations, and head models provides a rationale for basing protection against adverse effects on the PNS electrostimulation at intermediate frequencies. Future studies should revise the effects of higher frequencies that may affect the definition of stimulation thresholds that tend to be more compromised under the current results. Importantly, the results and methodology of this study could potentially contribute to the rationale and serve as input for the revision or updates of the intermediate frequency exposure guidelines.

ACKNOWLEDGMENT

The authors would like to IEEE/ICES/TC95/SC6 members for the valuable discussions during the conduction of this work.